\documentclass[prl,amsmath,amssymb,twocolumn,showpacs]{revtex4}
\usepackage{graphicx}
\usepackage{epsfig}
\usepackage{bm}

\begin{document}
\title{Spectroscopy of Valley Splitting in a Silicon/Silicon-Germanium
Two-Dimensional Electron Gas} 
\author{Srijit Goswami$^1$} 
\author{Mark Friesen$^1$}
\author{J. L. Truitt$^1$} 
\author{Charles Tahan$^2$} 
\author{L. J. Klein$^1$} 
\author{J. O. Chu$^3$}
\author{P. M. Mooney$^4$}
\author{D. W. van der Weide$^5$}
\author{S. N. Coppersmith$^1$} 
\author{Robert Joynt$^1$} 
\author{Mark A. Eriksson$^1$} 
\affiliation{$^1$Department of Physics, University of
Wisconsin, Madison, WI 53706} 
\affiliation{$^2$Cavendish Laboratory, Madingly Road, Cambridge CB3 OHE, U.K.}
\affiliation{$^3$IBM Research Division, T. J. Watson
Research Center, NY 10598}
\affiliation{$^4$Department of Physics, Simon Fraser University, Burnaby, BC V5A 1S6 Canada}
\affiliation{$^5$Department of Electrical and Computer Engineering, University of
Wisconsin, Madison, WI 53706} 

\begin{abstract}
The lifting of the two-fold degeneracy of the conduction valleys in a strained silicon 
quantum well is critical for spin quantum computing. Here, we obtain an accurate 
measurement of the splitting of the valley states in the low-field region of interest, 
using the microwave spectroscopy technique of electron valley resonance (EVR).  We 
compare our results with conventional methods, observing a linear magnetic field 
dependence of the valley splitting, and a strong low-field suppression, consistent 
with recent theory.  The resonance linewidth shows a marked enhancement above 
$T\simeq 300$~mK.

\end{abstract}


\maketitle

The term ``valley physics" 
refers to the study of degenerate valleys in the conduction band of an
indirect gap semiconductor such as silicon.  Valley physics has become a focal point in the
field of silicon spintronics and quantum information processing because of the couplings
between valley and spin states.  For instance, in the Kane quantum computer \cite{kane98},
the interactions between spins are strongly modulated by interference between the different
valleys \cite{koiller02}.  Similar concerns exist for spin qubits in a quantum
well \cite{eriksson04}.  

Although valley physics has emerged as an important field of study,
many important experimental questions remain unsettled,
due to the dearth of valley-sensitive measurement techniques, particularly for 
a two dimensional electron gas (2DEG).  
One example is the discrepency between theory and experiment for the
magnitude of the energy gap between the ground and excited valley states (the so-called valley
splitting).  While theory predicts that the valley splitting should be of order 1~meV for a  
silicon/silicon-germanium quantum well \cite{boykin04}, experimental measurements 
can be 10-100 times smaller \cite{weitz96,koester97,khrapai03,lai04,pudalov}.  
Typical detection techniques involve beating in Shubnikov-de Haas measurements or 
activation energy analyses.  These methods are difficult to apply with high precision, 
and they do not work well at the low fields of interest for quantum devices.  
Since valley splitting must be large enough to minimize 
excitations outside the qubit Hilbert space of a spin-based quantum computer \cite{friesen03},
it is crucial to understand the low-field valley physics, and to perform accurate low-field
measurements.

In this Letter, we apply the high precision microwave resonance techniques developed  
for spin excitation to the problem of valley splitting in a 2DEG.  Because of
the small number of electrons and the low-temperature requirements,
conventional microwave absorption spectroscopy techniques (which
require almost 10$^{12}$ electrons for an appreciable resonance
signal \cite{jiang01}) are difficult to employ in these structures.  On the other hand, 
transport measurements are naturally suited for probing electrical
characteristics of narrow channels containing relatively few carriers.
Previous studies have combined these techniques in the form of electrically detected 
electron spin resonance (ED-ESR), thereby enabling the measurement of
Zeeman splitting in both gallium arsenide 
\cite{jiang01,dobers88,dobers88b,vitkalov00,olshanetsky03}
and silicon \cite{graeff99,tyryshkin05} 2DEG structures.  
Here, we show that electronic transitions can also be driven between
the two lowest valley states using microwaves to achieve electrically detected 
electron valley resonance (ED-EVR).  The advantage of this technique is that it allows 
accurate and dense data acquisition over more than a decade of low magnetic fields.

The Si/SiGe heterostructures used in these experiments were grown by
ultrahigh vacuum chemical vapor deposition \cite{ismail95}.  In each case, the 2DEG
is located atop 80 $\textrm{\AA}$ of strained Si grown on a
strain-relaxed Si$_{0.7}$Ge$_{0.3}$ buffer layer.  The 2DEG is
separated from the phosphorus donors by 140 $\textrm{\AA}$ of
Si$_{0.7}$Ge$_{0.3}$.  The donors lie in a 140 $\textrm{\AA}$ layer of
Si$_{0.7}$Ge$_{0.3}$ with a 35 $\textrm{\AA}$ Si cap
at the surface.  Further details about the structure can be found in
reference \cite{klein04}.  Two 2DEGs ($S1$ and $S2$) were measured at 
0.25~K, obtaining the electron densities
$n=4.2\times 10^{11}\, \text{cm}^{-2}$ ($S1$) and 
$n=5.5\times 10^{11}\, \text{cm}^{-2}$ ($S2$), and the mobilities 
$50,000\, \text{cm}^{2}$/Vs ($S1$) and $180,000\, \text{cm}^{2}$/Vs ($S2$).  

\begin{figure}[t]
\begin{center}
\centerline{\includegraphics[width=2.7in]{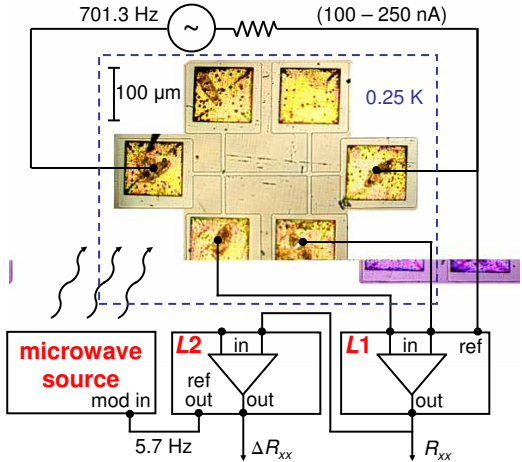}}
\caption{Schematic of the experimental setup, using a double lock-in
technique ($L_1$ and $L_2$).
\label{fig:schematic}}
\end{center}
\end{figure}

A schematic of the experimental set-up is shown in Fig.~\ref{fig:schematic}. A
double lock-in technique
is used to measure the change in resistance $\Delta R_{xx}$ of the 2DEG 
as a function of the perpendicular field
$B$, in the presence of microwaves. Lock-in 1
provides a bias current ranging from 100~nA to 250~nA, modulated at
701.3~Hz. Lock-in 2 is used to modulate the microwave amplitude
with 100\% modulation at 5.7~Hz. The output of Lock-in 1 is fed into
Lock-in 2, which measures $\Delta R_{xx}$.
Microwaves are produced by an HP83650A synthesizer, and are carried
down to the sample using a low loss coaxial line terminating about 5
cm from the surface of the sample in a loop antenna. The base of a
resonant cavity is replaced with a sample stage. 
The microwave power at the sample has a strong frequency
dependence because of the open cavity and the impedence mismatches along
the length of the coaxial line.  Because of this non-uniformity, a wide
range of powers (10 $\mu$W-10 mW) are used to ensure optimal
power delivery. The magnetic field is produced by a
superconducting magnet and all measurements are carried out in an
Oxford Instruments $^3$He cryostat with a base
temperature of 0.25~K.

The same experimental set-up can be used to detect both ESR and EVR
signals. Although we do not report on ESR here, we observe
typical resonances, with linewidths on the order of 5~G for $S1$ and 2~G for $S2$. The
EVR transition is slightly different than ESR because it
is not driven by magnetic fields.  (The two low-lying valley states 
are orthogonal and unaffected by the spin operator, causing the 
Zeeman transition matrix element to vanish.) 
However, the electric dipole transition is allowed by general symmetry considerations 
\cite{kleiner}, which also apply to the quantum well geometry.
We can estimate the magnitude of this valley excitation using a one-dimensional
tight binding (TB) method \cite{boykin04}.
Since the two valley states differ only along the $z$ direction (the
direction perpendicular to the quantum well), only the $E_z$ component
of the microwave field may induce transitions.  The resulting dipole matrix element
is rather small.  Nonetheless, it is the dominant transition mechanism.   
By further positioning the sample at $B$ or $E$ nodes in the resonant cavity, it may
be possible to drive valley or spin excitations selectively, although we do not perform
such experiments here.

\begin{figure}[t]
\begin{center}
\centerline{\includegraphics[width=2.5in]{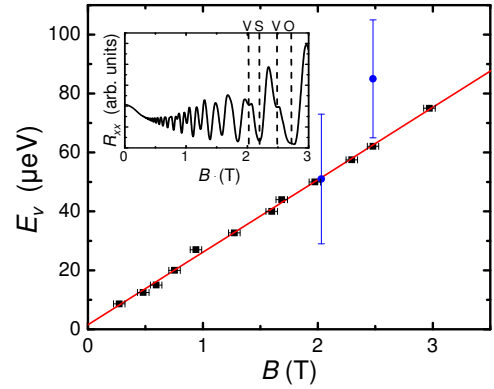}}
\caption{Valley splitting data for two different samples.  Results for sample $S1$ (squares) 
were obtained using EVR.  An activation energy analysis was used for sample $S2$ (circles).
The line represents a linear fit to the transport data for $S1$.  
Inset:  Shubnikov-de Haas data shows splitting of the orbital oscillations (O),
due to the lifting of spin (S) and valley (V) degeneracies.
\label{fig:linear}}
\end{center}
\end{figure}

Shubnikov-de Haas oscillations provide a rough estimate of the
valley splitting, as demonstrated in the inset of Fig.~\ref{fig:linear} for sample $S2$.
Both spin and valley features can be observed in the data \cite{lai04}. 
By increasing the magnetic field, we observe a sequential removal of the spin and
valley degeneracies, as indicated by the appearance of split spin and valley peaks.
In the main figure, we present a measurement of the valley splitting (circles), based on 
an activation energy analysis \cite{weitz96}.  The analysis provides results only at 
higher magnetic fields.  The larger error bars reflect uncertainties in the
fitting of the activation data, similar to previous experiments \cite{weitz96}.

\begin{figure}[t]
\begin{center}
\centerline{\includegraphics[width=2.9in]{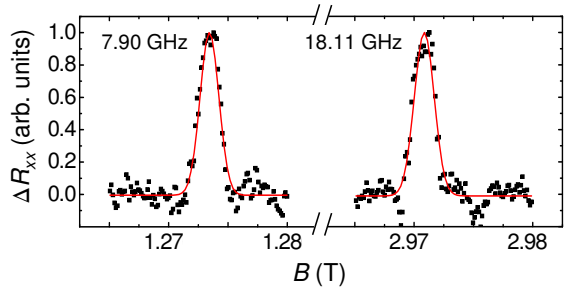}}
\caption{Electrically detected EVR signals at two different microwave
frequencies with Gaussian fits.
\label{fig:peaks}}
\end{center}
\end{figure}

In contrast, the electrically-detected EVR measurement provides narrow
error bars, and covers a wide range of 
magnetic fields.  Here, we obtain data from 0.27~T to 3~T on sample $S1$, with
narrow error bars throughout.  Some typical resonances are shown in Fig.~\ref{fig:peaks}.
To analyze the resonance features, we fit the data.  First, the background 
resistance is removed by fitting to a second degree polynomial away
from the main peak.  We find that gaussians provide the best representation of the 
individual peaks, with peak widths on the order of 20-25~G.  Typically, the 
resonance features account for about one part in $10^4$  
of the total resistance signal.  In Fig.~\ref{fig:peaks}, 
the the peak heights have been scaled to unity. 

The fitted peak positions are plotted
in Fig.~\ref{fig:linear} (squares), as a function of the perpendicular magnetic field. 
To estimate the error bars, we note that
the microwave power dependence of the resonance peak shows a shift
towards higher fields with increasing power.  The largest observed
shift is about 100~G over two orders of magnitude in the power. In the
figure, we determine our error bars using a more generous estimate of 500~G.
The resulting data are strikingly linear, with regression giving a slope of
$(24.7\pm 0.4)\,\, \mu \text{eV}/\text{T}$, and a $y$-intercept of
$E_v(0) = (1.5 \pm 0.6) \, \mu \text{eV}$.  A separate linear fit of the valley 
splitting data from $S2$ suggests a slightly larger slope for $E_v(B)$.
Although this fit has larger error bars than the EVR analysis, the results
appear consistent with theoretical expectations that the valley splitting should scale
with the 2DEG density as $E_v\propto n$ \cite{theory}.

\begin{figure}[t]
\begin{center}
\centerline{\includegraphics[width=2.5in]{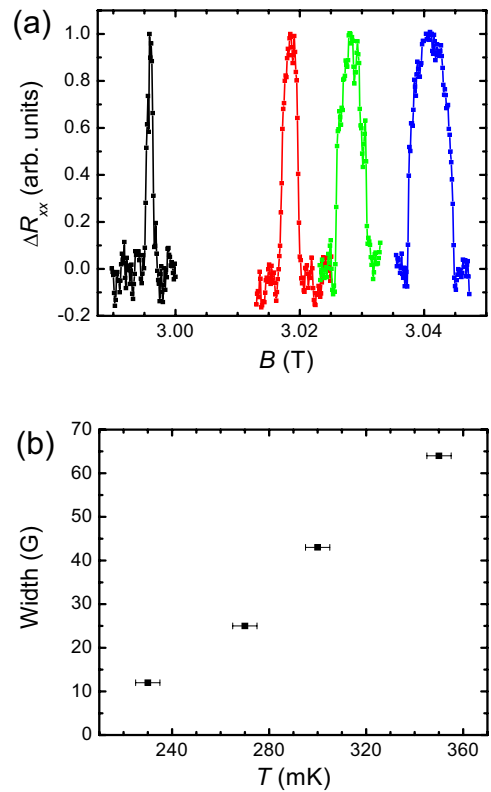}}
%\centerline{\includegraphics[width=2.6in]{EVR4bc.eps}}
\caption{Temperature dependence of the valley resonance.
(a) Normalized EVR signals from 230~mK (left) to 350~mK (right). 
(b) Temperature dependence of the corresponding fwhm linewidths.
\label{fig:temperature}}
\end{center}
\end{figure}

Figure~\ref{fig:temperature} shows the effect of temperature on the resonance peaks. 
We use the frequency (18.11~GHz) and magnetic field (3.0~T) that give the 
largest valley splitting.  Data sets are taken as the sample temperature is increased from 
0.23~K to 0.35~K and decreased back to 0.23~K at a slow rate, giving the sample enough time 
to equilibriate.  The error bars for Fig.~\ref{fig:temperature}(b) are obtained from the hysteresis in peak widths during the temperature cycle.  Over this narrow temperature 
range there is a sudden, rapid (seven-fold) increase in the linewidth.  
Resonances at higher temperatures are difficult to observe,  
although valley splitting can still be observed in the activation energy analysis, for
temperatures up to 1~K.  In addition to thermal broadening 
of the resonance, we also observe a small but reproducible increase in the
peak position of about 400~G, or 1.5\%, indicating a small thermal 
enhancement of the valley splitting.  

While the data in Fig.~\ref{fig:linear} are internally consistent, they give values
for valley splitting that are far smaller than the theoretical estimates.  
In the conventional theory of valley splitting, the degeneracy of the conduction valleys
is broken by the sharp confinement potential of the quantum well, which couples
the valleys in $k$ space
\cite{ohkawa77,sham79,AFS}.  Tight binding
and non-equilibrium Green's functions techniques obtain estimates of 0.1-1~meV for
the valley splitting, depending on the quantum well width and the internal electric 
field associated with modulation doping \cite{boykin04}.  It has been
suggested that the magnetic field dependence arises from
the enhancement of exchange coupling due to electron-electron 
interactions in the 2DEG \cite{ohkawa77,AFS,shkolnikov02}.
However, the many-body effect is expected to enhance the 
valley splitting, in contrast with the suppression observed in experiments.  
In addition, the linear dependence of $E_v(B)$ is 
inconsistent with the expected scaling for the many-body theory \cite{shkolnikov02}.  

Ando has proposed an alternative explanation
for the suppression of valley splitting \cite{AFS,ando79}.  In this picture,
the sharp confinement potential is still the cause of the valley splitting.  However,
one must also include the effects of substrate miscut and rough growth surfaces.  
Indeed, the commercial substrates used for SiGe heterostructures are often purposely miscut.
The samples used in this work were miscut at a $2^\circ$ angle.
Quantum wells grown on such substrates will be misaligned with respect to the 
crystallographic axes.  For rough surfaces, there will be an additional, locally varying 
misalignment.

A theory of valley splitting on a stepped quantum well is given 
in Ref.~\cite{theory}, based on effective mass theory.  A number of the
experimental features in Fig.~\ref{fig:linear} appear consistent with the 
theoretical predictions.
We now briefly discuss the theory, and its implications for our work.  The valley 
splitting can be expressed as a simple integral,
$E_v = 2\left|\int e^{-2ik_0 z} |F({\bm r})|^2 V_v({\bm r})d^3r \right|$, where $k_0$ is 
the position of the $z$ valley minimum, $F({\bm r})$ is the effective mass envelope
function, and $V_v({\bm r})$ is a valley coupling interaction, caused by the sharp
interface of the quantum well.  Each step in the quantum well makes a contribution to the
integral with the phase 
$e^{-2i k_0z_j}$, where $z_j$ is the interface position of the $j$th step.
Since the difference in the phase angles on consecutive steps is 
$2k_0b\simeq 0.85\pi$, where $b=1.35$~\AA\ is the atomic step height, the step  
contributions interfere destructively.
Thus, an electronic wavefunction covering many steps will have a
valley splitting that is strongly suppressed compared to the flat quantum wells of 
Ref.~\cite{boykin04}.
Since stepped surfaces are ubiquitous in conventional semiconductor heterostructures, so too
is the suppression of the valley splitting.  In a magnetic field, on the other hand,
the electron is confined to a finite number of
steps, thereby limiting the destructive interference.  For very large magnetic fields,
the electron may be confined to a single step.  In this limit, the valley splitting is 
restored, and $E_v$ achieves its theoretical upper bound.

The arguments given in Ref.~\cite{theory} for the strong suppression of $E_v$ at very 
low fields are plausible and consistent with our data.  The partial lifting of 
this suppression due to step disorder and other fluctuations is also 
plausible.  In Ref.~\cite{theory}, simulations were performed on disordered steps,
including bunched steps.  These obtain valley splitting results very
close, quantitatively, to the data of Fig.~\ref{fig:linear}.  However, the shape of the
$E_v(B)$ curves obtained from the simulations depends on the particular disorder model, making
a definite theoretical comparison difficult.  Most noteably, the simulation results do not 
appear completely linear, contrary to our experimental observations.  A particular
``plateau" model was suggested in that work, as an example of a disorder model
producing a linear $E_v(B)$.  However, experimental verification of such behavior
is not yet available.  On the other hand, the proposed model of envelope function 
oscillations at $B=0$ gives an estimate for the valley splitting which is 
very close to the extrapolated of value for $E_v(0)$, obtained from Fig.~\ref{fig:linear}.

There are other open questions in the valley splitting theory of Ref.~\cite{theory}.
The EVR experiments summarized in Figs.~2 and 3 exhibit sharp resonance peaks, 
indicating a well-defined valley splitting.  However, models involving disorder suggest 
that electrons can become localized in the valley splitting landscape.  In this picture, 
electrons may fill both shallow and deep pinning sites, consistent with a range of 
valley splittings.  One might therefore expect a broad resonance 
peak, in contrast with experimental observations.  It is tempting to 
attribute the marked thermal broadening of the
EVR peaks to the increased occupation
of higher energy states.  A more complete theory must also take into account the fact that
our electrical detection method is inherently dynamical. 

Finally, we discuss the importance of our results for quantum computing in a silicon
2DEG.  To minimize the excitation of the valley states in a spin-based
quantum computer, the
valley splitting should be much larger than the temperature.  For a system
cooled to 100~mK, a valley splitting of 100~$\mu$eV should be adequate.  
We have noted that a strongly confined electron will exhibit a larger valley splitting.
For the plateau  
model of Ref.~\cite{theory}, the linear dependence of $E_v(B)$ can be expressed through
$E_v= C/\pi R^2\theta^2$, where $\theta$ is the miscut angle, and
$R=\sqrt{2}l_B=\sqrt{2\hbar/|eB|}$ is the rms radius of the magnetically confined 
wavefunction.  We can extend this scaling to confinement produced by gate potentials.
For sample $S1$, with $\theta =2^\circ$, our experimental results lead to
$C=4.1\times 10^{-19}\,\text{eV}\,\text{m}^2$.  If we now take $R$ to be the  
quantum dot radius and require $E_v=100\, \mu$eV, 
we obtain the relation
$R\theta =36$~nm.  To attain $E_v=100\, \mu$eV on a $2^\circ$ miscut therefore
requires a fairly small dot radius of 18~nm.  However, for a $0.3^\circ$ miscut, a  
larger 120~nm dot can be used.

In conclusion, we have performed microwave spectroscopy
of the conduction valley states in a strained silicon 2DEG using
transport measurements.  Our method provides a measurement of the valley
splitting over a wide range of relatively low magnetic fields (0.3-3~T).  We also
obtain temperature dependent measurements showing a strong change in the valley splitting
behavior above 300~mK.  We compare our data to a recent theory of
valley splitting on a stepped substrate obtaining quantitative agreement, 
but leaving some open questions.  

\begin{acknowledgments}
We gratefully acknowledge conversations with R. Blick.
This work was supported by NSA and ARDA under ARO contract  
number W911NF-04-1-0389, and by the National Science Foundation through the ITR 
program (DMR-0325634) and the QuBIC program (EIA-0130400).
\end{acknowledgments}


\begin{thebibliography}{99}

\bibitem{kane98}
B. E. Kane, Nature (London) \textbf{393}, 133 (1998).

\bibitem{koiller02}
B. Koiller, X. Hu, and S. Das Sarma, \prl \textbf{88}, 027903 (2001).

\bibitem{eriksson04}
M. A. Eriksson, \textit{et al.}, Quant. Inform. Process. \textbf{3}, 133 (2004).

\bibitem{boykin04}
T. B. Boykin, \textit{et al.}, Appl. Phys. Lett. \textbf{84}, 115 (2004);
\prb \textbf{70}, 165325 (2004).

\bibitem{weitz96}
P. Weitz, \textit{et al.}, Surface Science \textbf{361/362}, 542 (1996).

\bibitem{koester97}
S. J. Koester, K. Ismail, and J. O. Chu, Semicond. Sci. Technol.
\textbf{12}, 384 (1997).

\bibitem{khrapai03}
V. S. Khrapai, A. A. Shashkin, and V. P. Dolgopolov, \prb \textbf{67}, 113305 (2003).

\bibitem{lai04}
K. Lai, \textit{et al.}, \prl \textbf{93}, 156805 (2004).

\bibitem{pudalov}
V. M. Pudalov \textit{et al.}, cond-mat/0104347.

\bibitem{friesen03}
M. Friesen, \textit{et al.}, \prb \textbf{67}, 121301 (2003).

\bibitem{jiang01}
H. W. Jiang and Eli Yablonovitch, \prb \textbf{64}, 041307 (2001).

\bibitem{dobers88}
M. Dobers, K. von Klitzing, and G. Weimann, \prb \textbf{38}, 5453 (1988).

\bibitem{dobers88b}
M. Dobers, K. von Klitzing, J. Schneider, G. Weimann, and K. Ploog,
\prl \textbf{61}, 1650 (1988).

\bibitem{vitkalov00}
S. A. Vitkalov, C. R. Bowers, J. L. Simmons, and J. L. Reno, \prb \textbf{61}, 5447 (2000).

\bibitem{olshanetsky03}
E. Olshanetsky \textit{et al.}, \prb \textbf{67}, 165325 (2003).

\bibitem{graeff99}
C. F. O. Graeff \textit{et al.}, \prb \textbf{59}, 13242 (1999).

\bibitem{tyryshkin05}
A. M. Tyryshkin, S. A. Lyon, W. Jantsch and F. Schaeffler, \prl \textbf{94}, 126802 (2005).

\bibitem{ismail95}
K. Ismail, M. Arafa and K. L. Saenger, Appl. Phys. Lett. \textbf{66}, 1077 (1995).

\bibitem{klein04}
L. J. Klein \textit{et al.}, Appl. Phys. Lett. \textbf{84}, 4047 (2004).

\bibitem{kleiner}
W. H. Kleiner and W. E. Krag, \prl \textbf{25}, 1490 (1970).

\bibitem{theory}
M. Friesen, M. A. Eriksson, and S. N. Coppersmith, unpublished.

\bibitem{ohkawa77}
F. J. Ohkawa and Y. Uemura, Journ. Phys. Soc. Japan, \textbf{43}, 925 (1977).

\bibitem{sham79}
L. J. Sham and M. Nakayama, \prb \textbf{20}, 734 (1979).

\bibitem{AFS}
T. Ando, A. B. Fowler, and F. Stern, Rev. Mod. Phys. \textbf{54}, 437 (1982).

\bibitem{shkolnikov02}
Y. P. Shkolnikov, E. P. De Poortere, E. Tutuc, and M. Shayegan,
\prl \textbf{89}, 226805 (2002).

\bibitem{ando79}
T. Ando, \prb \textbf{19}, 3089 (1979).

\end{thebibliography}
\end{document}